# Current State and Future Directions for Learning in Biological Recurrent Neural Networks: A Perspective Piece

**Luke Y. Prince**[1,2,*,†,‡] | **Roy Henha Eyono**[1,2,‡] | **Ellen Boven**[3,‡] | **Arna Ghosh**[1,2,‡] | **Joe Pemberton**[3,‡] | **Franz Scherr**[4,†] | **Claudia Clopath**[5] | **Rui Ponte Costa**[3,§] | **Wolfgang Maass**[4,§] | **Blake A. Richards**[1,2,6,§] | **Cristina Savin**[7,§] | **Katharina Anna Wilmes**[8,§]

[1]School of Computer Science, McGill University, Montreal, Canada

[2]Mila, Montreal, Canada

[3]Bristol Computational Neuroscience Unit, University of Bristol, United Kingdom

[4]Institute of Theoretical Computer Science, Graz University of Technology, Graz, Austria

[5]Bioengineering Department, Imperial College London, London, United Kingdom

[6]Department of Neurology and Neurosurgery, Montreal Neurological Institute, McGill University, Montreal, Canada

[7]Center for Neural Science, New York University, New York, USA

**Correspondence**
Roy Henha Eyono
Email: roy.eyono@mila.quebec

[*]Equally contributing authors.
[†]moderator
[‡]co-organizer
[§]panellist

We provide a brief review of the common assumptions about biological learning with findings from experimental neuroscience and contrast them with the efficiency of gradient-based learning in recurrent neural networks. The key issues discussed in this review include: synaptic plasticity, neural circuits, theory-experiment divide, and objective functions. We conclude with recommendations for both theoretical and experimental neuroscientists when designing new studies that could help bring clarity to these issues.[a]

**KEYWORDS**
recurrent neural networks, backpropagation through time,

> synaptic plasticity
>
> [a] see recording at https://www.youtube.com/watch?v=SErwiM693AI

# 1 | INTRODUCTION

Biological agents excel at learning tasks involving long-term temporal dependencies. One of the key challenges in learning such tasks is the problem of temporal credit assignment: reliably assigning importance to past neural states for error observed in present time. The tasks that animals have to solve on a daily basis depend on learning both short-term and long-term associations. Understanding how temporal credit assignment is implemented in biological recurrent neural networks (RNN) is important for understanding the brain [1].

The nervous systems of mammals, birds, and even many invertebrates are populated with recurrent networks, in which action potentials sent by neurons can often be returned via one or two hops to other local neurons. On a larger scale, whole brain regions form loops in long-range communication. These loops in communication present a challenge to learning, as credit assignment mechanisms must determine both when and where in the network a change must be made to improve future performance. In artificial RNNs, the credit assignment problem can be solved using backpropagation-through-time (BPTT) and an optimal choice of architecture [2]. However, the suite of operations and memory available to biological neurons are not considered sufficient to implement backpropagation-through-time. BPTT requires artificial neurons to constantly track the synaptic activity of neurons that they are not connected to, which is biologically problematic as it assumes perfect sequence recall [1]. Nevertheless, biological neural networks have a plethora of mechanisms for transmitting information amongst neurons across multiple spatiotemporal scales. Identifying these mechanisms, or alternatively, showing that biological networks have ways of sidestepping the temporal credit assignment problem, is one of the key challenges for neuroscientists over the next decade.

Additionally, animals possess an aptitude for learning throughout their lifetimes in which they can maintain a strong generalized ability over a diverse range of tasks. However, modern artificial recurrent neural networks lose their generalized ability partly due to catastrophic forgetting [3]. Hence one of the reasons why they struggle to match their biological counterparts and are typically specialized over a narrow task domain. As a result, the network's performance on previously learned tasks is diluted.

In this position piece, we present the perspectives of several experts in theoretical neuroscience to reconcile the gap between biological and artificial learning in recurrent neural networks. This will not be an exhaustive review of the literature on synaptic plasticity, temporal credit assignment, recurrent neural circuits, or continual learning, however we will guide the reader on reviews of these topics. This piece will be organised in four sections that broadly summarise the main points of discussion in the workshop:

1. Bio-plausible approaches for solving temporal credit assignment using synaptic plasticity
2. The role of neural circuits and architecture in temporal credit assignment
3. The nature of objective functions in the recurrent circuits of the brain
4. What can experiments tell us: the limitations in comparing computational models to experimental evidence

At the beginning of each section, we will outline, in broad terms, what our panellists had agreed upon, followed by where their opinions diverged. We will end this piece with practical recommendations for theoretical and experimental research on how we can collectively help to resolve these issues over the coming years.

## 2 | BIO-PLAUSIBLE APPROACHES FOR SOLVING TEMPORAL CREDIT ASSIGNMENT

Our panellists were in agreement that temporal credit assignment must in some way be reliant on synaptic plasticity at recurrent synapses (see Citri and Malenka [4] and Magee and Grienberger [5] for reviews on synaptic plasticity). Furthermore there was a general consensus that recurrent plasticity is as important during early development (e.g., when learning to walk) as it is in adulthood (e.g., when learning to be a better dancer).

Experimentally, we observe that synaptic strengths are highly plastic [6]. In early development the conditions for changing a synapse are far less strict than in adulthood. As an animal grows older, synaptic plasticity becomes more gated and dependent on the presence of additional factors such as neuromodulators (see Pawlak et al. [7] for review on neuromodulators) or the release of inhibition. Furthermore, synapses in sensory areas tend to be less plastic than in higher-order areas [8, 9], although plasticity at sensory recurrent synapses can still be observed under the right conditions [10, 8, 9, 11].

To date there have been few successful attempts at implementing a biologically-plausible temporal credit assignment mechanism that can solve non-trivial tasks, at the efficiency of backpropagation-through-time in deep recurrent networks. The most successful approaches approximate gradients using a combination of neuromodulators and local estimates of the interactions that neurons have on each other over time [12, 13]. Nevertheless, there is still a gap when it comes to their efficacy as a general-purpose replacement for back-propagation-through-time [14]. It is still unknown whether their efficacy is restricted to particular problem domains. It is plausible that different brain regions leverage different strategies for learning the temporal structure of sensory data and that the specific solutions offered by these algorithms are restricted to particular tasks for particular brain regions, in particular species. The algorithms discussed here do not, by any means, incorporate every possible method to spread local and non-local information about network activity across time (e.g., dendritic spikes [15], disinhibition [16], acetylcholine [17], synaptic cooperativity [18], extrasynaptic transmission [19], presynaptic inhibition [20]). As such the range of possible solutions to temporal credit assignment in biological RNNs remains vastly underexplored.

Additionally, the reason for the success of biological recurrent networks in lifelong learning remains a mystery. There are a few solutions to this problem that have been proposed in artificial neural networks (e.g. elastic weight consolidation [21], hypernetworks [22] and intelligent synapses [23]). A commonality amongst some of these approaches is in increasing the complexity of the synapse models beyond scalar $w_{ij}$ terms [24]. Indeed, synapses are themselves highly nonlinear and have multiple gating and memory mechanisms that could allow for rapid storage, recall, and preservation of parameter sets important for a particular task, but that does not contribute to or suffer from interference with other task solutions [25].

Overall, the panellists acknowledge that current computational models of temporal credit assignment falls short on the capabilities of recurrent synapses in neural circuitry. It is however unclear if understanding the bio-complexity requires to reexamine the traditional set of synaptic learning rules (i.e. Hebbian plasticity) or rather to be more inclusive on what constitutes bio-plausibility when considering the current computational building blocks.

# 3 | THE ROLE OF NEURAL CIRCUITS AND ARCHITECTURE IN TEMPORAL CREDIT ASSIGNMENT: ARE WE FOCUSING TOO MUCH ON SYNAPTIC PLASTICITY?

It's well established that some connectivity motifs and microcircuits are conserved across species and individuals (e.g., neocortical columns, the hippocampal-entorhinal formation, cortico-cerebellar loops, and striatal networks). Furthermore, a certain degree of functional specialization consistently emerges across brain regions in terms of hierarchies within visual, somatosensory, motor, or auditory cortices. This degree of conservation indicates that much of the information required to generate these architectures are stored genetically, as a consequence of animal evolution. While panellists agreed that it is likely too extreme to say (in mammals at least) that learning relies on pre-wired recurrent circuits (as in echo state networks or liquid state machines [26]), there are clearly constraints imposed by the genetic code that may offer useful inductive biases for self-organization and learning.

Do these conserved microcircuit architectures support efficient temporal credit assignment, or potentially offer a strong inductive bias that need not be updated based on experience? Recent studies have shown that synaptic weight changes are selective and respect the prior architectural traits. One study mapped the key features of the cortical microcircuit onto state of the art gated-RNNs (e.g. LSTMs or GRUs) in machine learning [27]. Furthermore, two studies investigating the neural mechanisms underlying the transformation of sensory neuronal activity to associated motor actions have revealed that learning the auditory discrimination task preferentially potentiated synapses corresponding to high or low frequencies, depending on the reward structure [28, 29]. Furthermore, it was observed in a follow-up study that synaptic weights did not change if the task structure was altered after the animal had learned the old task [28]. Although the aforementioned experiments targeted projections from one region to another, they favour the conjecture that microcircuit architectures offer a strong inductive bias with limited plasticity in recurrent connections.

Nevertheless, it is difficult to show experimental evidence that supports whether fixed recurrent circuits play a part or not at all. All panellists agreed that designing an experiment, along the lines of those mentioned above, and studying synaptic plasticity in task-relevant recurrent connections during learning will offer a better insight to resolve this issue. Furthermore, the role of plasticity in recurrent circuits would be demystified if we can assess learning deficits when these specific synaptic weight changes are disrupted. Notably, a significant challenge could be designing a task that requires the agent to rely on good temporal credit assignment. It has often been observed in computational studies that performance in several associative learning tasks can be explained using a liquid state machine with fixed dynamics and plasticity only at feedforward readout weights. Arguably, typical behavioural tasks investigated by neuroscientists may not need a deep credit assignment mechanism [30].

# 4 | WHAT'S THE BRAIN'S OBJECTIVE FUNCTION?

Our panellists agree that good credit assignment likely requires synaptic plasticity to be at least a rough approximation of gradient descent along some objective function. Exploiting statistical temporal regularities is critical for survival, and even basic sensory tuning requires some form of credit assignment.

The nature of the objective function is largely unknown. Since animals are rarely given precise external feedback, it is likely that learning is driven by intrinsic signals, often referred to as self-supervised learning. One prevalent hypothesis is that a major driver of learning is prediction, in which microcircuits are trained to predict their own future activity and propagate prediction errors or surprise signals to neighbouring microcircuits [31, 32]. In doing so, neurons learn to group together external features with similar characteristics, the fundamental basis for a world-model with predictable causal interactions.

While this form of predictive learning may be important in the brain, it is likely that the brain learns over multiple objectives in different brain regions that may cooperate or compete with each other. Indeed there is evidence that brain areas can compete for control (e.g., habit-based or goal-directed brain areas competing for action selection), and that brain areas can cooperate (feedback association cortices guide plasticity in sensory cortices). Multi-objective training is widespread in deep neural networks, with examples of adversarial (e.g. Variational Autoencoders [33] and Generative Adversarial Networks [34]) and co-operative (e.g. Decoupled Neural Interfaces [35] and Siamese Networks [36]) training. However, there is disagreement if from an experimental point of view competition versus coorperation is a testable hypothesis. This raises the question of how can we evaluate theory in experimental work.

## 5 | WHAT CAN EXPERIMENTS TELL US: THE LIMITATIONS IN COMPARING COMPUTATIONAL MODELS TO EXPERIMENTAL EVIDENCE

A hard question for both theoretical and experimental neuroscientists is *How do you know when plasticity is due to learning?* Fundamentally, the panellists agree that it is experimentally challenging to monitor a large number of synapses over time and relate their changes to behaviour. Synaptic strengths can change for a variety of reasons, e.g., homeostatic mechanisms that try to maintain constant degrees of neural activity at a single-cell and population level over time (see [37, 38, 39] for homeostatic mechanisms that increase or decrease weights for reasons other than any obvious learning). Furthermore, memory traces and neuronal assemblies that are formed and linked to behaviour continue changing over time.

To date, no technology exists that allows massive, direct, chronic monitoring of synaptic plasticity at a high temporal resolution (see Humeau and Choquet [40] and Tsutsumi and Hayashi-Takagi [41] for reviews on SoTA for measuring functional changes in synaptic strength related to learning). However, Tony Zador contributed to our discussion by outlining technological developments that could soon allow large-scale monitoring of some forms of plasticity [42, 43] (See video). The proposed technique relies on tagging glutamate receptors that are inserted into the post-synaptic membrane which can then be visualized post-mortem and associated with its parent neuron. In conjunction with a behavioural task, this allows observation of changes in synaptic strength that occurred over the course of learning and post-mortem *in-vitro* experiments could be used to interrogate the plasticity rules at synapses where changes have been recorded. While this does represent an advance over current experimental approaches, it is still far from observing the dynamics of synaptic strengths during the course of learning. Although it is yet to be realised, such a technology would likely have a transformational effect on the field in terms of our understanding of learning in neural circuits,

However, the lack of tools for tracking synaptic weights over time across a network has not discouraged researchers from inferring plasticity mechanisms from observations of somatic or dendritic activity [44]. As such, it should be possible to glean some insights from experiments using techniques that are currently available. For example, Gillon et al (2021) demonstrate that unexpected event signals in individual neurons and distal apical dendrites of the visual cortex can differentially predict subsequent changes in responses to expected and unexpected stimuli over days [45]. At the same time, it can be shown in artificial neural networks that it is possible to decode which update rules are being employed based solely on node activations [46, 47]. This shows that it may be possible to infer some information about synaptic dynamics from neural dynamics, if model assumptions about how synaptic plasticity relates to neural dynamics and learning can be clearly stated.

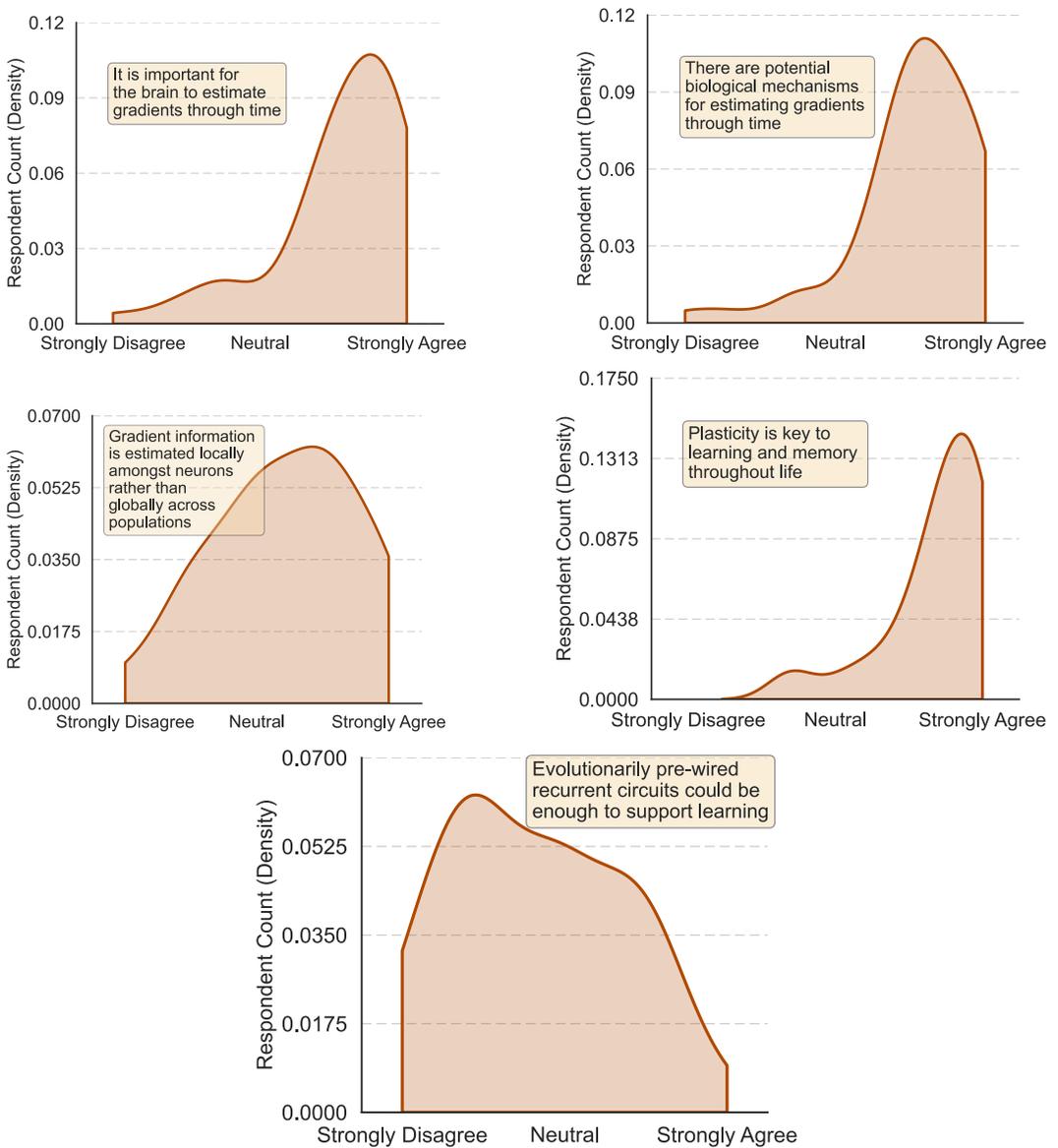

**FIGURE 1** Density plots illustrating the audience sentiment towards a series of questions regarding synaptic plasticity, gradient computation and evolutionary neural circuits. 58 respondents had participated in this questionnaire.

## 6 | CONCLUSION

Much of the difficulty in resolving how temporal credit assignment is implemented in biological recurrent networks comes from limitations in linking changes in synapses to changes in behaviour. At the same time, biologically-inspired solutions for learning in artificial recurrent neural circuits is vastly underexplored. What are the specific tasks, circuits, and species we can use to think about these issues?

> **Recommendation Box 1** A priority for theoretical neuroscientists should be to expand the model space:
>
> - We need to propose *concrete* and *experimentally testable* approximate credit assignment mechanisms that respect the space and time complexity of operations in neural circuits.
> - Of crucial importance, a proposed algorithm has to solve a non-trivial task *in silico*.
> - We should analyse the commonalities and differences across these different computational approximations.

An effective grouping of different theories will make experimental predictions more feasible to test, and allow us to understand when and where these proposed credit assignment mechanisms are best deployed.

> **Recommendation Box 2** Experimental neuroscientists should prioritise designing behavioural analogues for common machine learning tasks that exploit an animal's natural behaviour:
>
> - We need to design tasks that are complicated enough to require non-trivial credit assignment that cannot be achieved with a shallow feedforward network, i.e., should be non-linearly separable, yet still experimentally feasible.
> - We should concurrently record neural activity of somatic, and ideally, dendritic activity through imaging techniques to support the efforts of computational neuroscientists.

Good examples of non-trivial tasks for credit assignment include motor control tasks, delayed non-match to sample tasks, and path integration tasks during natural foraging or fear behaviours. Similarly, recordings from somatic and dendritic activity could potentially be directly compared with the learned representations and observed dynamics in deep artificial neural networks, where the architecture, optimizer, and cost functions are fully specified. This data may help to reveal the most plausible temporal credit assignment strategies used by the brain.

We concluded our workshop with an audience poll in order to evaluate the audience's takeaway and alignment to the panelists' opinions. Figure 1 indicates the distribution of responses that were submitted.

## Acknowledgements

The authors thank Surya Ganguli, Tony Zador, Guillaume Lajoie, and Alexandre Payeur for their active participation with panellists during the workshop. We would also like to thank all other audience members from around the world who participated on the day in the chat, and everybody else who has watched the replay since it was posted online. Finally, we'd like to that Megan Peters, Eric deWitt, Nicolas Kriegeskorte, and the rest of the team from CCN who made the workshop possible.